\documentclass[%
 reprint,
superscriptaddress,
 amsmath,
 amssymb, 
 aps,  
]{revtex4-2}

\usepackage{graphicx}
\graphicspath{{plots/fig1/}{plots/fig2/}{plots/fig3/}{PDF_output/}{Final_plots/}{figures/}}
\usepackage{comment}
\usepackage{dcolumn}
\usepackage{bm}
\usepackage{hyperref}

\usepackage{xfrac}
\usepackage{siunitx}
\usepackage[T1]{fontenc} 
\usepackage[dvipsnames]{xcolor}
\usepackage{ulem}
\usepackage{tikz}
\usepackage{upgreek}
\usepackage{graphicx}
\usepackage{hyperref}

\colorlet{ForestGreen}{Black}
\begin{document}

\preprint{APS/123-QED}
\title{Readout of strongly coupled NV center-pair spin states with deep neural networks}

\author{Matthew Joliffe}
 \affiliation{3rd Institute of Physics, IQST, and Research Center SCoPE, University of Stuttgart, Stuttgart, Germany}
  
 \author{Vadim Vorobyov}
\email[]{v.vorobyov@pi3.uni-stuttgart.de}
 \affiliation{3rd Institute of Physics, IQST, and Research Center SCoPE, University of Stuttgart, Stuttgart, Germany}  

\author{J\"org Wrachtrup}
 \affiliation{3rd Institute of Physics, IQST, and Research Center SCoPE, University of Stuttgart, Stuttgart, Germany}
 \affiliation{Max Planck Institute for solid state physics, Stuttgart, Germany}

\begin{abstract}
Optically addressable electron spin clusters are of interest for quantum computation, simulation and sensing. However, with interaction length scales of a few tens of nanometers in the strong coupling regime, they are unresolved in conventional confocal microscopy, making individual readout problematic.
Here we show that when using a single shot readout technique, collective states of the combined register space become accessible.  
By using spin to charge conversion of the defects we draw the connection between the intricate photon count statistics with spin state tomography using deep neural networks. 
This approach is particularly versatile with further scaling the number of constituent spins in a cluster due to complexity of the analytical treatment.
We perform a proof of concept measurement of the correlated classical signal, paving the way for using our technique in realistic applications. 
\end{abstract}
\maketitle

\section{Introduction}
Spin defects in solids have emerged as promising quantum systems which serve as versatile and powerful test models for numerous quantum applications \cite{Awschalom_2018}. 
Among recent demonstrations, quantum simulations of spin hamiltonians \cite{randall2021many}, building of quantum registers \cite{Bradley2019,vorobyov2021quantum}, and distributed quantum sensing \cite{rovny2022nanoscale} showcase the usage of single color centers. Recent advances in material fabrication \cite{haruyama2019triple, ji2024correlated} motivate the extension of the platform beyond a single color center. 
One method to scale-up the local spin register is utilizing strongly coupled arrays of quantum dots, spin qubits, or color centers \cite{broome2018two, huxter2024multiplexed, oakes2023fast, nurizzo2023complete, he2019two, dolde2013room}.
The necessity of strong coupling ($g>\gamma$) dictates emitters to be within $d \le \SI{20}{nm}$ ($g \ge \SI{10}{kHz}$) and hence beyond the resolution of a conventional confocal microscopy ($\delta x > \SI{300}{nm}$).
This has the implication that readout couples to several systems at once, complicating the readout, which is a key prerequisite for operation of a spin register. 

The techniques available for single defects in a confocal spot operation \cite{zahedian2023readout}, nuclear \cite{neumann2010single, neumann2010single}, and electron spin \cite{robledo2011high, rosenthal2024single} readout could potentially be combined with super-resolution techniques, such as electrical assisted readout \cite{siyushev2019photoelectrical} with nanoscale contracts, optical super-resolution imaging \cite{jaskula2017superresolution} or spectral multiplexing \cite{Bersin_2019} at low temperatures. 
However, most of these techniques cannot be broadly and flexibly applied to color centers. For example, application of super-resolution techniques are limited when dealing with nanostructures, due to spectral diffusion and broadening at low temperatures. 
Hence, efficient readout remains limited to a single NV center per confocal spot \cite{shields2015efficient, zhang2021high}.

Here we explore an alternative method where we exploit the joint addressability of defects and harness the dynamics of their response signals. 
When multiple defects couple to the detection channel, a readout signal forms a multiplexed telegraph-like process, showing stochastic switching behaviour between multiple discrete values. 
Upon calibration of a system's transition parameters, spin to charge conversion methods enable each defect to be readout with high efficiency \cite{aslam2013photo, shields2015efficient,  zahedian2023readout}.
Spin to charge conversion at room temperatures operate with a contrast similar to the population of the metastable singlet state.
 After the initial excitation and the calibration process, extracting the precise spin states requires multiple measurements. 
Recalibration due to parameters drifting and the growing complexity with increased number of defects make this cumbersome.
In contrast, machine learning provides a viable approach to reading out a many qubit system. 
Here we perform spin to charge conversion combined with machine learning assisted statistical analysis for efficient readout of a diffraction unresolved spin cluster spaced by $d \sim 10$ nm. 
We perform a calibration of the readout using uncorrelated and correlated states, and perform state tomography of the system. 
Our results pave the way towards correlation sensing and scalable quantum hardware based on dipolar coupled spin arrays of color centers.


\section{Results}
\begin{figure*}[ht]
  \centering
  \includegraphics{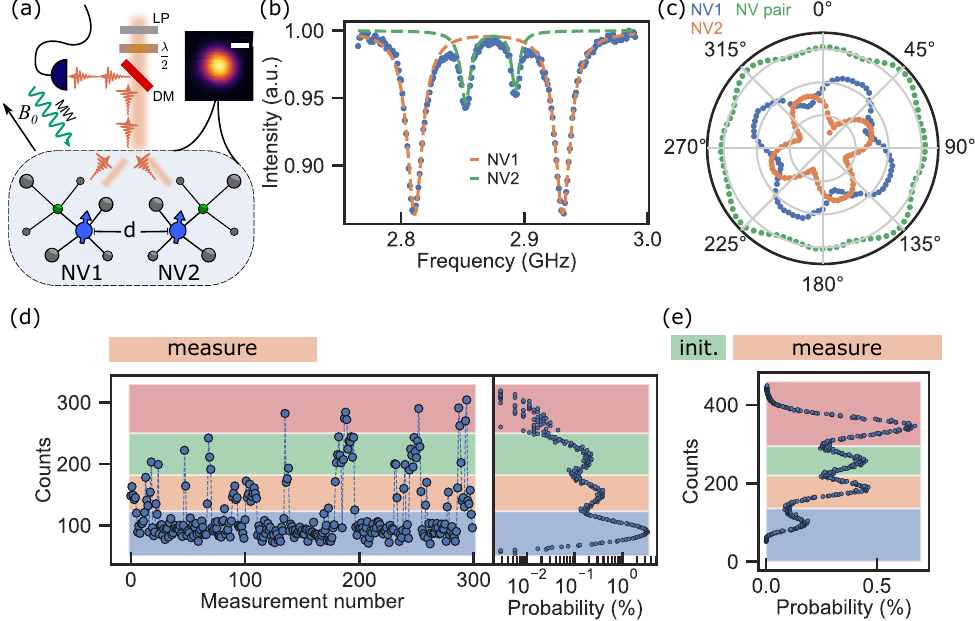}
  \caption{
Charge state readout of the nitrogen vacancy (NV) pair system 
a) Schematic image of a differently oriented NV pair addressed with a confocal microscope. 
White bar on the confocal image is 200 nm in length. NV centers addressed with optical illumination and microwave frequency. 
b) Optically detected magnetic resonance of the NV-pair system. 
c) Polarization intensity plot. 
Blue, orange – intensity of NVs of different orientation as the excitation laser polarization is rotated. Green – combined intensity of the NV pair
d) Timetrace of photon countrate under continuous weak illumination with a 594 nm laser.
e) Histogram of timetrace of 594 readout with a short 520 nm laser for charge state initialization.
}
  \label{fig:figure1}
\end{figure*}

In our experiments we use an NV pair of multiple orientations created using the ion implantation reported in \cite{jakobi2016efficient}. 
The NV pair system Fig. \ref{fig:figure1}a. is in the strong coupled regime $g \gg \Gamma_2$. 
The individual NVs are within the diffraction limited spot that makes individual readout a challenge.
Our test system is comprised of two NV centers with two different crystal lattice orientations. 
The different orientations can therefore allow for individual addressability of the NV centers due to the different magnetic field alignment and hence transition frequencies. 
For example, optically detected magnetic resonance of the NV centers shows multiplexed signal corresponding to each of the NV centers Fig. \ref{fig:figure1}b.
A challenge of a full state readout is to assign the signal to individual centers and perform a readout of the spin correlations. 
Here we exploit the alignment of the excitation laser polarization along one of the optical dipoles of NV1 (see Fig. \ref{fig:figure1}c). 
Consequently, we increase the emission intensity of the selected NV1 orientation and decrease the intensity of the other orientation for NV2. 
This provides a way to create distinguishability between the two emitters' brightnesses which can be harnessed for demultiplexing the readout in a single shot readout setting, e.g. of the defect charge state. 
The NV center has two main charge states, NV$^-$ and NV$^0$, with their zero phonon line emission at wavelengths of 637 nm and 575 nm, respectively. 
By choosing a wavelength between these values ($\sim 594$ nm), we can efficiently excite only the NV$^-$ charge state. 
Fig. \ref{fig:figure1}c shows a continuous readout of the two NV system. 
As the system jumps between the four possible charge states, the intensity varies. 
Fig. \ref{fig:figure1}d shows a histogram of the photon counting statistics of many charge state readouts. 
Distinction of 4 peaks in the histogram signals contribution of two two-level systems contributing to the observation with four possible combined states: NV$_1^-$ NV$_2^-$, NV$_1^0$ NV$_2^0$, NV$_1^-$ NV$_2^0$, NV$_1^0$ NV$_2^-$. 
When the NV pair system has its charge state prepared with a short 520 nm pulse prior to the charge state estimation, the overal probability of having the useful double negative state increases, reaching $\sim 40\%$ (Fig. \ref{fig:figure1}e).
To maximize the distinguishability of the histogram peaks, we apply the polarization alignment such that NV1 is roughly twice as bright as NV2. The distribution of each charge state is expected to be a Poisson distribution, but these are slightly altered due to the charge state often switching mid readout \cite{shields2015efficient,  zahedian2023readout}.

\begin{figure*}[ht]
  \centering
  \includegraphics{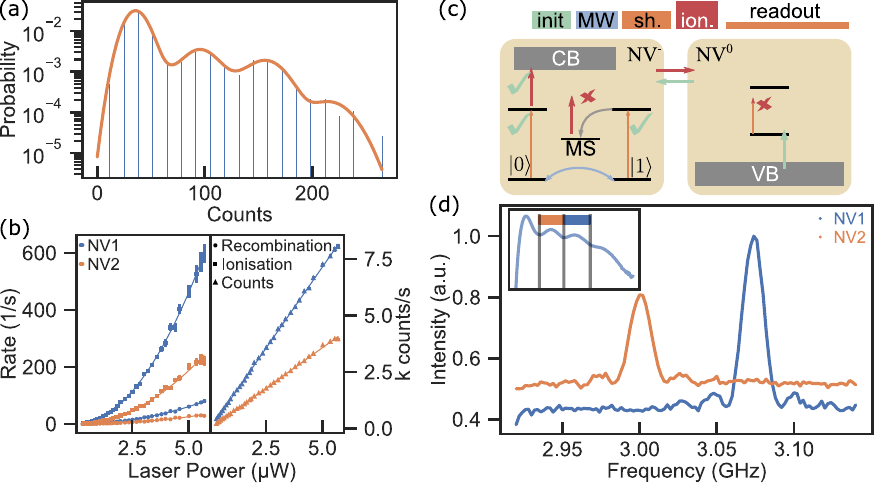}
  \caption{Charge switching dynamics of an NV pair and their spin to charge conversion for readout demultiplexing.
a) Histogram of counts of the NV pair under continuous 594 excitation. Fit with a two independent switching emitter model (see text).
b) Left - charge state switching rates as a function of laser power. Ionization is a two-photon process and thus increases quadratically with laser power. Right - emission intensity increases linearly in the low-power regime.
c) Schematic of the spin to charge conversion protocol. A 594 pulse cycles spin states between the ground and excited states. The spin 1 state has a significant probability
of decaying through the metastable singlet state, where it is protected from a high power 638 pulse, while the spin 0 state is ionized with high probability. 
The charge state
can then be read out with a lower power 594 pulse. The spin and charge state are then reset with a 520 pulse.
d) Demultiplexed readout of optically detected magnetic resonance. A histogram is measured for each step of an ODMR measurement with spin to charge conversion (inset). The relative intensities of
the middle two peaks are plotted as a function of probed frequency, and the two NV centers are read out individually and simultaneously.
}
  \label{fig:figure2}
\end{figure*}

Due to the significant probability of charge state switching during a measurement, care must be taken when trying to assign the initial charge state at the time of readout. 
To account for the intricate charge state dynamics in photon distributions, we adapt the photon counting distribution derived for a single emitter and assume independence of their charge state switching events and rates: 
\begin{equation}
p(n,k_1,k_2) = \sum_{i=0}^n p_1(N=i,k_1) p_2(N=n-i,k_2),
\label{eq:charge}
\end{equation}
where the $p_i(N,k_1)$ photon counting probability distributions conditioned on initial state $k$, derived in \cite{shields2015efficient,zahedian2023readout} for single emitters. 
 Fig. \ref{fig:figure2}a shows the histogram of a measurement with continuous charge state readout. 
The model eq. \ref{eq:charge} is able to fit the obtained distribution. 
As the model depends on the intrinsic charge switching and photon emission rates, after performing a series of measurements at various laser intensities we estimated the ionization and recombination rates versus the applied 594 nm laser illumination (Fig. \ref{fig:figure2}b). 
This allows for the laser power and readout time to be optimized numerically for further readout use \cite{zahedian2023readout}. 
\begin{figure*}[ht]
  \centering
  \includegraphics{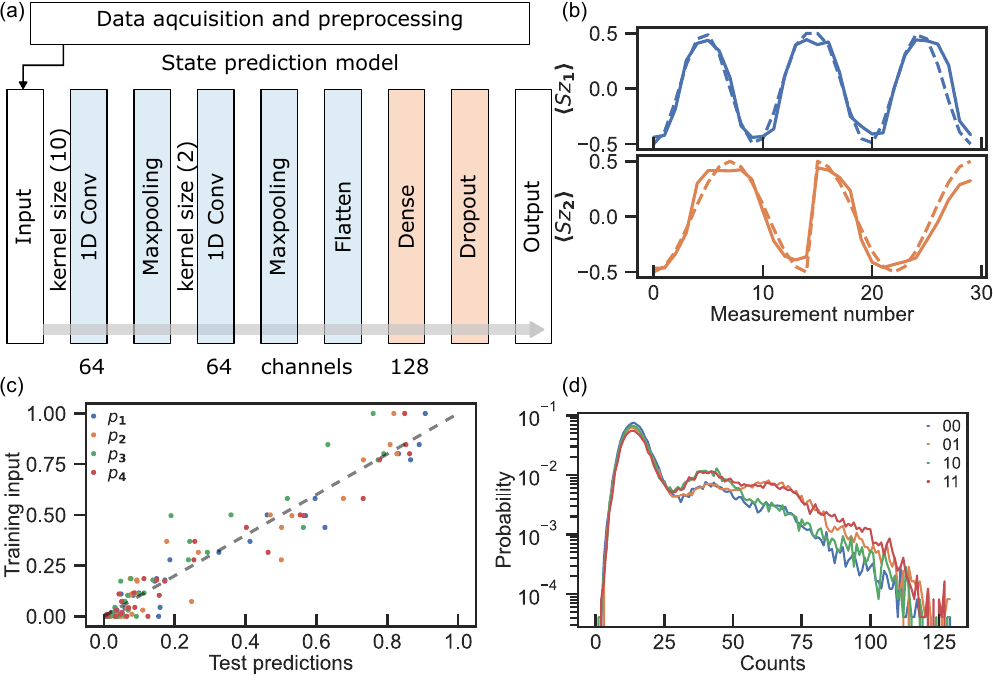}
  \caption{ \textbf{(a)} Neural network model for robust state readout a) overview of the model used in the experiment. 
  \textbf{(b)} Known states are prepared and the corresponding histograms measured. These are fed into the ML model where features are extracted and the model attempts to learn relationships between the shape of the histogram and the probabilities it corresponds to.
  It then predicts the probabilities, and these are compared to the actual values. The model updates its internal weightings and continues this process trying to minimize the loss function between its predictions and the true values. The loss function used is the mean square error with an additional L2 regularization penalty that encourages lower weightings.
\textbf{(c)} Comparison between the training data input and the predictions on the test set. The line y=x represents perfect predictions
\textbf{(d)} The photon counting histograms of the basis states for spin projections of individual NV centers correspondingly 00,01,10,11}
  \label{fig:figure3}
\end{figure*}

Next we use the ability to efficiently estimate the charge state and individual microwave control of the two NV centers to realize an efficient individual spin readout. 
To this end we map the spin of the centers to a long lived charge state. 
At room temperature, we cannot selectively ionize the spin states directly but we can use the meta stable state to achieve this.
The metastable single state (MS) provides protection from the defect getting ionized Fig. \ref{fig:figure2}c
This, along with the differential decay probabilities through the MS, we can ionize to the NV$^0$ state. 
The full protocol runs as following. 
A short 594 nm laser excitation pulse excites both spin states. 
The $m_s=0$ state is spin conserving and will cycle between the 0 ground and excited states. The $m_s=1$ state has a significant probability of decaying through the long-lived metastable singlet state. 
A short intense 638 nm laser pulse will then ionise the $m_s = 0$ state, leaving the MS untouched. 
The probability of the $m_s=1$ state being in the MS at the ionization pulse is ~30\%, similar to the ODMR contrast. 
The charge state can then be read out with a low power 594 nm laser pulse. 
Fig. \ref{fig:figure2}d demonstrates the combination of spin to charge conversion with charge state readout and individual spin control. 
 A pulsed ODMR measurement is performed using spin to charge conversion and charge state readout. For each frequency in the measurement, a histogram is acquired (inset in the \ref{fig:figure2}d). 
 First we digitize and count the number of events in the second and third peaks of the histogram (inset). 
As the microwave frequency is swept, the occurrences of these events vary, representing effectively the spin flip probability for both NV centers in a demultiplexed manner. 
As this clearly shows the ability for individual spin readout, the quantitative analysis and calibration for precise tomography of the defects is still cumbersome. 
This is due to complicated propagation dynamics of the spin states into the final charge state distribution and photon counting statistics, and depends significantly on such parameters as laser intensity, laser polarization and microscope alignment. 

\begin{figure*}[ht]
  \centering
  \includegraphics{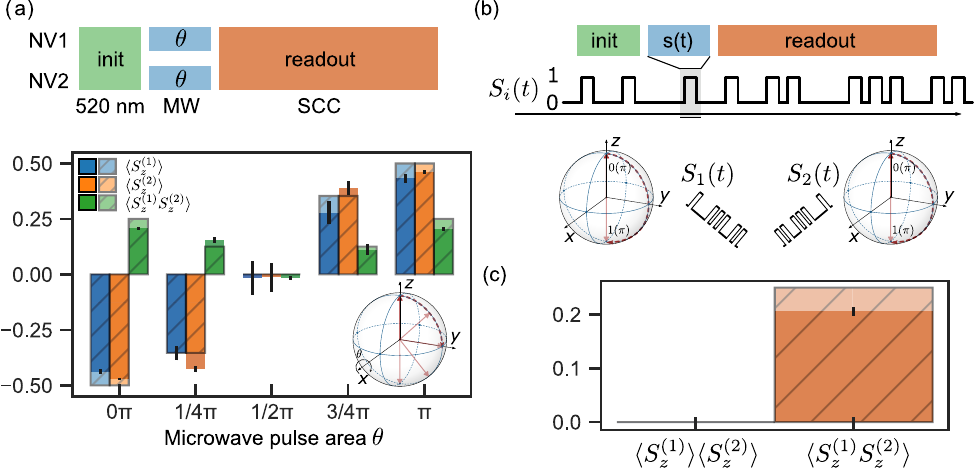}
  \caption{Tomography of the states and sensing. 
a) NV center pair prepared into the 00 state and then a fractional pulse is applied to both NVs and measured across many measurements. The histograms were fed into an already trained model and the expectation values obtained.
b) Quantum sensing of correlated zero mean signal with NV center signal. 
c) Reconstructed product of single spins and correlation of the spin values due to correlated signal (noise). 
}
  \label{fig:figure4}
\end{figure*}
Additionally, the fitting method of the obtained photon statistics distribution for extraction of occupancies of the charge state is complicated due the large parameter space and numerical complexity of the model. 
This is also poorly scalable to systems with more than two emitters.  
Thus, here we find it natural to introduce a neural network based machine learning approach for reading out complex spin states with the model presented in Fig. \ref{fig:figure3}a.
A normalized photon count histogram of a type shown in Fig. \ref{fig:figure2}a is fed in as an input vector to the model.
Following convolution and max-pooling layers are placed for feature extraction of the histograms.
Fully connected dense layers placed afterwards are aimed at pattern learning and map the extracted features of the distributions to the probabilities of the initial states. 
An output layer yields model's estimation of the probabilities. 
During training, the predictions are compared to the actual values and the model updates its weightings using a gradient method.
Using the  \texttt{Tensorflow} python package \cite{tensorflow}, we we able to train a model to learn the spin states based on the photon histogram dataset at the inputs. 
To this end, known states are prepared via Rabi-like oscillations and a training histogram is measured for each state presented in Fig. \ref{fig:figure3}b. 
 A separate test histogram is also measured for each of these states. 
After model training we perform cross-validation with the test dataset, which is shown in  Fig. \ref{fig:figure3}c.
It is important that the training and test data are kept separate to ensure that model doesn't overfit and its predictions can be generalized.
Fig. \ref{fig:figure3}d shows the basis state histograms. 
It is evident that the histograms have a lot of overlap and a bias in the basis states representation, for example the first peak corresponding to the "00" charge state always has the predominant contribution to the signal. 
The features related to the differences between the states "0-","-0" are related to exchange in the second and third peak height. 

\begin{figure*}[ht]
  \centering
  \includegraphics{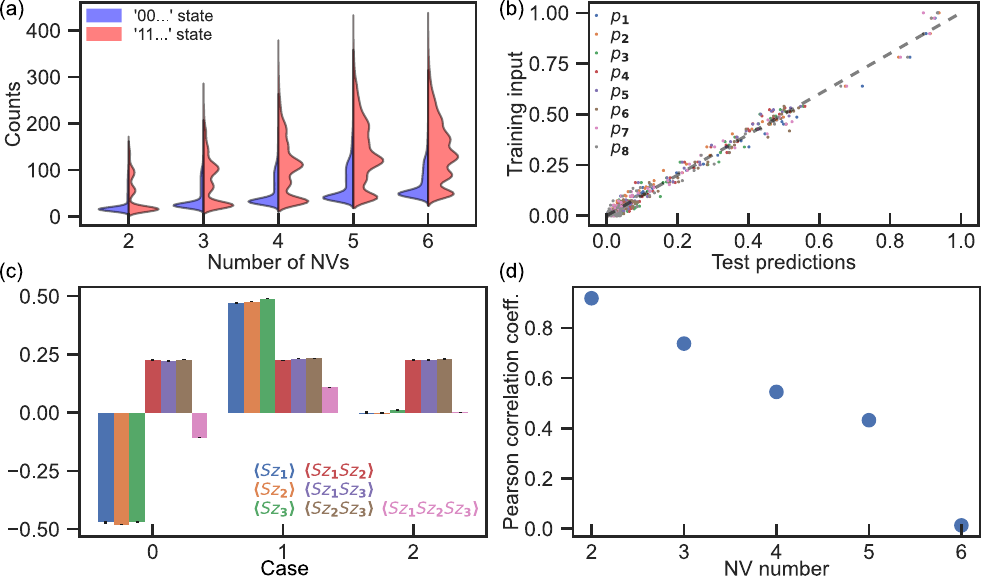}
  \caption{Scaling of the method to larger cluster sizes (numerical simulation) 
a) Simulated photon histograms for various number of NV centres for $|000....\rangle$ and $|111...\rangle$ states
b) Regression plot for training of the NN for an idealized three NV center system
c) State tomography for three NV centres analogous to Fig. \ref{fig:figure4}. Case "0" is pulse area $0\cdot\pi$, "1" is $1\cdot\pi$, and "2" is 50\% statistical mixture of cases 0 and 1. 
d) Scaling of Pearson coefficient for training of the model as a function of NV cluster size
}
  \label{fig:figure5}
\end{figure*}

Having trained the model we apply it to perform tomography and sensing using the NV center pair. 
Fig. \ref{fig:figure4}a shows the testing of the model on known states. 
Both NV center states are prepared in the negative charge with $m_s = 0$ state, then a rotation pulse with area $\theta = \{0,  1/4, 1/2, 3/4, 1\}  \times \pi$ are applied to both NV centres. 
For each of the final states 64 histograms were measured and were fed into the model to get their spin probabilities. 
The predictions for the 64 measurements were averaged to obtain the plotted values with the standard deviation as the error. 
After that, the expectation values for the operators $\langle S_z^{(1)} \rangle$, $\langle S_z^{(2)}\rangle$ and $\langle S_z^{(1)}S_z^{(2)}\rangle$ were determined. 
As expected, the parity measurement $\langle S_z^{(1)}S_z^{(2)}\rangle$ stays positive for all of rotation angles, paving the way for efficient quantum sensing of correlations at the nanoscale.
However, an interesting outcome is the result of the rotation to $\pi/2$ angle. It shows vanished correlations upon measurement in the $S_z$ basis, even though the spins were prepared to the same state, due to the probabilistic nature of the projective measurements which erases the created correlation. 
Next we show the sensing of correlated signals.
We perform a measurement where we apply simultaneously to both NVs correlated signal consisting of either an idle or an ideal $\pi$ rotation mixed in random sequence.
The Fig. \ref{fig:figure4}b shows the schematic of the sequence execution applied for both sensors. 
The results of the measurements are shown in the Fig. \ref{fig:figure4}c. Since the individual responses $\langle S_z^{(1,2)}\rangle$ are $\pm1/2$ upon $0(1)\pi$ pulses, the average expectation value of their product yields zero.
The correlated signal, in turn, persists, yielding a way to separate the two \cite{rovny2022nanoscale}.

\section{Discussion}
In this work we studied spin to charge assisted readout of electron spins of an NV center pair.
However, this method could be extended to larger number of centers in a cluster.
In Fig. \ref{fig:figure5} we show how the performance of the network scales with the number of defects.
Interestingly, we found that for the network to distinguish between the individual spins, it is not neccessary for them to differ in their fluorescence rates.
It is sufficient that they have a distinct property, e.g. charge switching rates or fluorescence rates.
This becomes crucial when the defects have same dipole orientation, as with increased number of defects, where some of them will inevitably have the same crystal orientation.
We simulated photon statistics histogram produced by defect clusters of various sizes shown in Fig. \ref{fig:figure5}a.
For three emitters, as demonstrated previously, we show the correlation plot in Fig. \ref{fig:figure5}b and the simulated correlated sensing protocols in Fig. \ref{fig:figure5}c.
However, with increasing cluster size, there are naturally less and less distinguishable properties of the emitters and the Pearson score for the networks drops close to 0 for a cluster of size 6, which shows the limits of the technique.

In conclusion, spin to charge conversion plus charge state readout allows for a novel way to read out two dipolar coupled NV centers simultaneously.
Different orientations of NV centers within the diamond lattice allow for distinguishability via emission intensity which can be tuned with excitation polarization.
As quantum systems get larger, readout becomes more challenging.
Here we show machine learning is a viable way to readout systems consisting of multiple defects.
A system of many emitters with switching charge states means significantly overlapping counts that would pose problems for traditional readout methods.
Machine learning allows us to bypass the need to describe the system exactly by having the model learn the patterns that represent different spin states.
Our method could be further applied for large clusters up to five systems.
Applications to other technological platforms are straightforward.
Due to the simplicity of the charge state readout, it could be straightforwardly applied to scanning probe setups, such as an NV-AFM \cite{huxter2024multiplexed}. 
Applications to quantum registers \cite{dolde2013room,joas2024high}, covariance magnetometry sensing \cite{rovny2022nanoscale}, and quantum repeaters are within reach.
Furthermore, charge readout of spin clusters posses additional insights into preferentially aligned spin clusters, where the Zeeman shift yields a degenerate magnetic resonance spectrum. 
 This technique could be further applied to a more general class of high fidelity readout, such as resonant readout, and repetitive readout. 

\acknowledgments
We acknowledge support from the European Commission through the QuantERA project InQuRe (Grant agreements No. 731473, and 101017733), the German ministry of education and research for the project InQuRe (BMBF, Grant agreement No. 16KIS1639K), support from the European Commission for the Quantum Technology Flagship project QIA (Grant agreements No. 101080128, and 101102140), the German ministry of education and research for the project QR.X (BMBF, Grant agreement No. 16KISQ013) and Baden-Württemberg Stiftung for the project SPOC (Grant agreement No. QT-6) and QC4BW as well as KQC4BW,  project Spinning (BMBF, Grant agreement No. 13N16219) and the German Research Foundation (DFG, Grant agreement No. GRK2642, FOR 2724).

\appendix
\section{NN training Loss function}
Machine learning model created with TensorFlow 2.17.0 in Python 3.11.7. The model was structured such that it has an input layer that accepts data in the format of the histogram. The input layer feeds the data to a convolutional layer that extracts features from the histogram. The data is put into a max-pooling layer that reduces its spatial dimension, abstracting the data, and focuses on the most prominent features. From there the data goes into the dense layer that performs high level reasoning, learns to associate certain features and patterns with spin states. From there, the output layer makes a final prediction of the quantum state.
The evolution of the training loss function is given in Fig. \ref{fig:figure6}a.

\section{Setup}
\begin{figure*}[ht]
  \centering
  \includegraphics{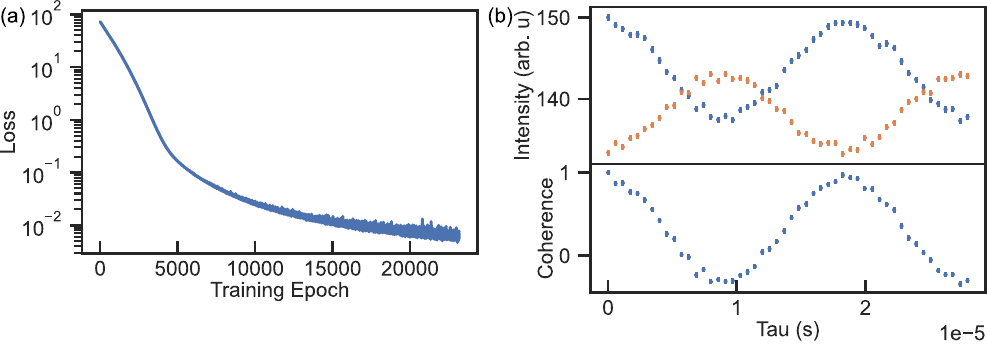}
  \caption{
a) Loss function evolution upon training of the model for the experimental case of NV center pair
b) Double electron-electoron resonance (DEER) NV-NV spectroscopy, revealing a $g\approx 2 \pi \cdot 50$ kHz coupling strength between the two color centers.
}
  \label{fig:figure6}
\end{figure*}

Sample implanted as in \cite{jakobi2016efficient}. Proximity confirmed with dipolar interaction measurements to be within $d\sim 10$ nm ($g\approx 2\pi \cdot  50$ kHz) inferred from NV-NV DEER spectroscopy depicted on Fig. \ref{fig:figure6}b.
Experiments were carried out with a home built confocal microscope. Homebuilt 520 green and 638 red diode lasers were combined with a 594 orange Huebner photonics Cobolt Mambo laser by coupling them into the same polarization maintaining Thorlabs P5-405BPM-FC-2 fibre. The polarization of all three lasers was set after outcoupling with Thorlabs LPVISC100 linear polariser, WPQ10M-588 quarter wave plate, WPH10M-588 half wave plate mounted on Thorlabs KPRM1E/M rotational stage to control the direction of polarization.
Keysight M8190A 12 GSa/s arbitrary waveform generator was used for MW pulses for NV spin control. Emitted photons detected with Perkin-Elmer SPCM-AQRH single photon counting device and counted with a TimeTagger 20. A PulseStreamer 8/2 was used for coordinating the three lasers, the AWG, and the TimeTagger.

%


\end{document}